# SCIENTIFIC REPORTS

**OPEN**

# Triplet Cooper pairs induced in diffusive *s*-wave superconductors interfaced with strongly spin-polarized magnetic insulators or half-metallic ferromagnets

Jabir Ali Ouassou[1], Avradeep Pal[2], Mark Blamire[2], Matthias Eschrig[3] & Jacob Linder[1]

Interfacing superconductors with strongly spin-polarized magnetic materials opens the possibility to discover new spintronic devices in which spin-triplet Cooper pairs play a key role. Motivated by the recent derivation of spin-polarized quasiclassical boundary conditions capable of describing such a scenario in the diffusive limit, we consider the emergent physics in hybrid structures comprised of a conventional s-wave superconductor (*e.g.* Nb, Al) and either strongly spin-polarized ferromagnetic insulators (e.g. EuO, GdN) or halfmetallic ferromagnets (e.g. $CrO_2$, LCMO). In contrast to most previous works, we focus on how the superconductor itself is influenced by the proximity effect, and how the generated triplet Cooper pairs manifest themselves in the self-consistently computed density of states (DOS) and the superconducting critical temperature $T_c$. We provide a comprehensive treatment of how the superconductor and its properties are affected by the triplet pairs, demonstrating that our theory can reproduce the recent observation of an unusually large zero-energy peak in a superconductor interfaced with a half-metal, which even exceeds the normal-state DOS. We also discuss the recent observation of a large superconducting spin-valve effect with a $T_c$ change ~1 K in superconductor/half-metal structures, in which case our results indicate that the experiment cannot be explained fully by a long-ranged triplet proximity effect.

Combining materials with different types of quantum order can result in new quantum phenomena at their interface. One example is the interaction between superconducting and magnetic materials[1, 2], which besides its interesting fundamental physics has spawned the field of superconducting spintronics[3], and could lead to novel cryogenic spin-based applications.

Recently, several experimental works have been carried out on superconductors interfaced to strongly spin-polarized magnetic materials. The latter include both ferromagnetic insulators such as EuO or GdN[4, 5], with spin-polarizations ranging up to 90%, and half-metallic ferromagnets such as $CrO_2$ and $La_{2/3}Ca_{1/3}MnO_3$ (LCMO)[6–8]. In ref. 8, STM-measurements were performed on the superconducting side of a NbN/LCMO bilayer, and revealed an unusually large zero-energy peak in the density of states (DOS) which, surprisingly, exceeded even the normal-state DOS. Such a peak, often taken as a hallmark signature of odd-frequency pairing[9, 10], was also observed recently in Nb/Ho bilayers in ref. 11, albeit with a reduced magnitude. Moreover, resistance measurements probing the superconducting critical temperature $T_c$ in MoGe/Ni/Cu/$CrO_2$ multilayers revealed a change in $T_c$ of up to 1 K; this was attributed to the generation of long-ranged triplet pairs when the relative magnetization between ferromagnetic Ni and half-metallic $CrO_2$ was changed from parallel to perpendicular[7]. It would be of high interest to understand and model the findings in these experiments, yet such an endeavour is complicated by the fact that there up to recently has existed no convenient framework allowing for the study of strongly spin-polarized magnetic materials in contact with superconductors in the experimentally relevant diffusive regime of transport.

[1]Department of Physics, NTNU, Norwegian University of Science and Technology, N-7491, Trondheim, Norway. [2]Department of Materials Science, University of Cambridge, Cambridge, CB3 0FS, United Kingdom. [3]Department of Physics, Royal Holloway, University of London, Surrey, TW20 0EX, United Kingdom. Correspondence and requests for materials should be addressed to J.A.O. (email: jabir.a.ouassou@ntnu.no)

Motivated by this, we here present a solution of the quasiclassical Usadel equation[12] with arbitrarily strongly spin-polarized magnetic regions and obtain the DOS and $T_c$, using the generally valid spin-dependent boundary conditions derived in ref. [13]. We have applied this framework on superconductors interfaced to strongly spin-polarized ferromagnetic insulators and half-metallic ferromagnets, solving the equations selfconsistently in order to study the manifestation of triplet Cooper pairs induced in the superconductor. While previous works have considered the case of strong spin-polarization in the ballistic limit[14–24], we here present results valid for the diffusive regime of transport. We show that our theory is able to reproduce an unusually strong zero-energy peak, exceeding the normal-state value, induced in a superconductor as seen experimentally in ref. [8]. Moreover, we compute the $T_c$ shift when the magnetization in a spin-valve S/F/N/HM multilayer is rotated, and discuss the results in the context of the experiment described in ref. [7]. Our results indicate that the experimental measurements cannot be fully explained by a long-ranged triplet proximity effect, suggesting that some different physical mechanism may also be at play. We clarify the difference in length-scale for the inverse proximity effect in a superconductor and the length-scale for which a spin-valve effect occurs.

## Theory

**Quasiclassical theory.** In this paper, we employ the quasiclassical theory of superconductivity[12, 25, 26] to describe diffusive hybrid structures in equilibrium. With this approach, the main objective is to calculate the quasiclassical retarded propagator $\hat{\underline{g}}$ as a function of quasiparticle energy $\epsilon$ and position $z$, where the $z$-axis is along the junction direction. The propagator may then be used to calculate various physical observables of interest, such as the density of states, tunneling currents, and superconducting critical temperature. We use a hat to denote that the propagator has a 2 × 2 matrix structure in Nambu space, an underline to indicate a 2 × 2 matrix structure in spin space, and that we use the normalization convention $\hat{\underline{g}}^2 = 1$. The quasiclassical propagator can be calculated from the Usadel diffusion equation[12],

$$iD\,\partial_z(\hat{\underline{g}}\,\partial_z\hat{\underline{g}}) = \hat{\underline{U}}, \tag{1}$$

where $D$ is the diffusion constant, and $\hat{\underline{U}}$ is a material-dependent matrix potential that incorporates the effects of various self-energies and scattering processes. We will later generalize eq. (1) to strong ferromagnets, where the diffusion constants become spin-dependent. In superconductor/normal-metal hybrid structures, the matrix potential takes the form[12, 26]

$$\hat{\underline{U}} = [(\epsilon + i\eta)\hat{\tau}_3 + \hat{\underline{\Delta}}, \hat{\underline{g}}], \tag{2}$$

where $\epsilon$ is the quasiparticle energy, $\eta$ mimics an inelastic scattering rate, $\hat{\tau}_3 = \text{diag}(+1, -1)$ is the third Pauli matrix in Nambu space, and the superconducting order parameter $\Delta(z)$ is embedded in the antidiagonal matrix $\hat{\underline{\Delta}} = \text{antidiag}(+\Delta, -\Delta, +\Delta^*, -\Delta^*)$. Note that we follow the convention where sums and products of dimensionally incompatible matrices should be resolved by taking Kronecker products with identity matrices. For instance, in the above equation, $\hat{\tau}_3$ lacks an explicit structure in spin space, and should therefore implicitly be interpreted as $\hat{\tau}_3 \otimes \underline{\sigma}_0$, where $\underline{\sigma}_0 = \text{diag}(+1, +1)$ is the identity matrix in spin space.

The above equations must also be accompanied by the appropriate boundary conditions,

$$G_L L_L\left(\hat{\underline{g}}_L \partial_z \hat{\underline{g}}_L\right) = G_R L_R\left(\hat{\underline{g}}_R \partial_z \hat{\underline{g}}_R\right) = \hat{\underline{I}}, \tag{3}$$

where the subscripts indicate whether the quantities correspond the left or right side of the interface, $G_j = \sigma_j A/L_j$ is the bulk conductance of material $j$, $L_j$ is the material length, $A$ is the cross-sectional area of the interface, $\sigma_j$ is the intrinsic conductivity in the non-superconducting state, and $\hat{\underline{I}}$ is the matrix current[27–29] at the interface. In general, the matrix current depends on the propagators at both sides of the interface, as well as the physical properties of the interface itself. The simplest case is when the interface has a relatively low transparency and no spin-active properties, in which case the matrix current is given by the Kupriyanov–Lukichev tunneling equation[30]

$$2\hat{\underline{I}} = G_0\left[\hat{\underline{g}}_L, \hat{\underline{g}}_R\right], \tag{4}$$

where $\hat{\underline{g}}_L$ and $\hat{\underline{g}}_R$ are the propagators at the left and right sides of the interface, respectively, and $G_0 \ll G$ is the conductance of the interface. How to calculate the matrix current at spin-active interfaces will be discussed in the following sections.

In practice, when solving the equations above, it is convenient to use the Riccati parametrization of the propagator[14, 31–33],

$$\hat{\underline{g}} = \begin{pmatrix} \underline{N} & \\ & -\tilde{\underline{N}} \end{pmatrix}\begin{pmatrix} 1 + \underline{\gamma}\,\tilde{\underline{\gamma}} & 2\underline{\gamma} \\ 2\tilde{\underline{\gamma}} & 1 + \tilde{\underline{\gamma}}\,\underline{\gamma} \end{pmatrix}, \tag{5}$$

where tilde conjugation $\tilde{\underline{\gamma}}(z, \epsilon) = \underline{\gamma}^*(z, -\epsilon)$ is defined as a combination of complex conjugation $i \mapsto -i$ and energy $\epsilon \mapsto -\epsilon$, and the normalization matrices are defined as $\underline{N} = (1 - \underline{\gamma}\tilde{\underline{\gamma}})^{-1}$ and $\tilde{\underline{N}} = (1 - \tilde{\underline{\gamma}}\underline{\gamma})^{-1}$. Mathematically, this parametrization automatically satisfies the normalization condition $\hat{\underline{g}}^2 = 1$, and enforces the particle-hole symmetries of the propagator. The Riccati parameters $\underline{\gamma}$ and $\tilde{\underline{\gamma}}$ are also single-valued and bounded, and the parametrization is numerically stable relative to alternatives like *e.g.* the $\theta$-parametrization. Using the definitions of $\underline{N}$ and $\underline{\tilde{N}}$ in terms of $\underline{\gamma}$ and $\underline{\tilde{\gamma}}$, as well as the easily derivable identities $\underline{N}\underline{\gamma} = \underline{\gamma}\underline{\tilde{N}}$ and $\underline{\tilde{N}}\underline{\tilde{\gamma}} = \underline{\tilde{\gamma}}\underline{N}$, it can be shown that eqs (1) and (3) can be Riccati parametrized as

$$\partial_z^2 \underline{\gamma} = (2iD\underline{N})^{-1}(\underline{U}_{12} - \underline{U}_{11}\underline{\gamma}) - 2(\partial_z\underline{\gamma})\underline{\tilde{N}}\ \underline{\tilde{\gamma}}(\partial_z\underline{\gamma}),  \quad (6)$$

$$\partial_z \underline{\gamma} = (2GL\underline{N})^{-1}(\underline{I}_{12} - \underline{I}_{11}\underline{\gamma}), \quad (7)$$

where the notation $\underline{U}_{\tau\tau'}$ and $\underline{I}_{\tau\tau'}$ refer to the $(\tau, \tau')$ components in Nambu space of the matrix potential $\hat{U}$ and matrix current $\hat{I}$. The corresponding equations for $\underline{\tilde{\gamma}}$ can be found by tilde conjugation of the equations above. Together, the differential equations for $\gamma$ and $\tilde{\gamma}$ form a boundary value problem that can be solved numerically as long as we know the matrix potential and current.

While the equations above are sufficient to solve for the propagator of the system, these equations implicitly depend on the superconducting order parameter $\Delta(z)$ through eq. (2). We therefore need an equation which relates this order parameter to the propagator in order to find a selfconsistent solution. In equilibrium, the appropriate selfconsistency equation can be written[34]

$$\Delta(z) = \frac{1}{2}N_0\lambda \int_0^{\Delta_0 \cosh(1/N_0\lambda)} d\epsilon [f_s(z, \epsilon) - f_s(z, -\epsilon)]\tanh(\epsilon/2T), \quad (8)$$

where $f_s = (f_{12} - f_{21})/2$ is the singlet component of the anomalous propagator $\underline{f} = [\hat{g}]_{12}$, $N_0$ is the density of states per spin at the Fermi level, $\lambda$ is the BCS coupling constant, is the zero-temperature gap of a bulk superconductor, $T$ is the temperature of the superconductor, and $T_c$ is the critical temperature of a bulk superconductor. The above equation can be written in terms of the Riccati parameters using the equations $\underline{f} = 2\underline{N}\underline{\gamma}$ and $f_s(-\epsilon) = \tilde{f}_s^*(\epsilon)$. If we furthermore divide the equation by $\Delta_0$, and use the approximations $\cosh(1/N_0\lambda) \cong \exp(1/N_0\lambda)/2$ and $\Delta_0/T_c \cong \pi/e^c$ where $c$ is the Euler–Mascheroni constant, we obtain

$$\Delta(z)/\Delta_0 = \frac{1}{2}N_0\lambda \int_0^{\exp(1/N_0\lambda)/2} d(\epsilon/\Delta_0)[(\underline{N}\underline{\gamma})_{12} - (\underline{N}\underline{\gamma})_{21} - (\underline{\tilde{N}}\underline{\tilde{\gamma}})_{12}^* + (\underline{\tilde{N}}\underline{\tilde{\gamma}})_{21}^*]$$
$$\times \tanh\left(\frac{\pi}{2e^c}\frac{\epsilon/\Delta_0}{T/T_c}\right), \quad (9)$$

where all the Riccati matrices $\underline{N}$, $\underline{\tilde{N}}$, $\underline{\gamma}$, $\underline{\tilde{\gamma}}$ are functions of position $z$ and quasiparticle energy $\epsilon$. Note that the approximations above are only valid in the weak-coupling regime $N_0\lambda \ll 1$. In practice, $N_0\lambda \leq 1/4$ is sufficient to make the results insensitive to the cutoff, and we set $N_0\lambda = 1/5$. This result is expressed in terms of only the Riccati matrices $\underline{N}$, $\underline{\tilde{N}}$, $\underline{\gamma}$, $\underline{\tilde{\gamma}}$ and dimensionless quantities $\Delta/\Delta_0$, $\epsilon/\Delta_0$, $T/T_c$, $N_0\lambda$, making this version of the equation better suited for numerics than the equivalent eq. (8).

**Spin-active tunneling interfaces (1st order in $\varphi_n$ and $T_n$).** In the case of low-transparency spin-active junctions where the spin-mixing is weak, the matrix current may be written[13, 35, 36]

$$2\hat{I} = G_0[\hat{g}_L, \hat{g}_R] + G_1[\hat{g}_L, \hat{m}\hat{g}_R\hat{m}] + G_{MR}[\hat{g}_L, \{\hat{g}_R, \hat{m}\}] - iG_\varphi[\hat{g}_L, \hat{m}_L], \quad (10)$$

where the magnetization matrix $\hat{m} = \text{diag}(\boldsymbol{m}\cdot\boldsymbol{\sigma}, \boldsymbol{m}\cdot\boldsymbol{\sigma}^*)$, $\boldsymbol{m}$ is a unit vector that describes the interface magnetization, $\boldsymbol{\sigma}$ is the Pauli vector, and $\hat{g}_L$ and $\hat{g}_R$ are the propagator at the left and right sides of the interface, respectively. Note this version of the matrix current equation is for the left side of the interface; a similar equation for the other side of the interface can be found by letting $\hat{I} \mapsto -\hat{I}$ and L $\leftrightarrow$ R. Note that there are two different magnetization matrices $\hat{m}$, $\hat{m}_L$ in the equation: $\hat{m}$ refers to the average magnetization felt by a quasiparticle *transmitted* through the interface, while $\hat{m}_L$ refers to the magnetization felt by a *reflected* quasiparticle. If there is an interfacial magnetic misalignment, these two magnetizations will in general be different, and this may cause long-range triplet generation. The interface conductances in the equation above can be written[13]

$$G_0 = G_Q\sum_{n=1}^{N} T_n\left(1 + \sqrt{1 - P_n^2}\right), \quad G_1 = G_Q\sum_{n=1}^{N} T_n\left(1 - \sqrt{1 - P_n^2}\right), \quad G_{MR} = G_Q\sum_{n=1}^{N} T_nP_n, \quad G_\varphi = G_Q\sum_{n=1}^{N} 2\varphi_n, \quad (11)$$

where $T_n$, $P_n$, $\varphi_n$ are respectively the transmission probability, spin-polarization, and spin-mixing angle associated with each scattering channel $n$. The quantity $G_Q$ in the equations above is the conductance quantum $e^2/\pi$ (in units with $\hbar = 1$), while we interpret $G_0$ as the tunneling conductance, $G_1$ as a depairing term, $G_{MR}$ as a magnetoresistive term, and $G_\varphi$ as the spin-mixing term. Note that in this context, the polarization is defined as $P_n \equiv [T_{n\uparrow} - T_{n\downarrow}]/[T_{n\uparrow} + T_{n\downarrow}]$, where $T_{n\sigma}$ are the spin-dependent transmission probabilities, and $\sigma$ is the spin of a quasiparticle as measured along the quantization axis $\boldsymbol{m}$. In other words, the polarization determines how many spin-up vs. spin-down particles are transmitted through the spin-active interface for each transmissive conductance channel $n$. Note that these equations can be used with arbitrary interface polarizations $P \in [-1, +1]$, but only remain valid as long as the transmission probabilities $T_n$ and spin-mixing angles $\varphi_n$ are small. In general, the number of channels contributing to $G_\varphi$ can be different from the number of channels contributing to $\{G_0, G_1, G_{MR}\}$ since channels that are purely reflecting can contribute to the former. If we assume that all scattering channels have the same polarization $P$, then $G_1$ and $G_{MR}$ can be calculated straight from the polarization $P$ and tunneling conductance $G_0$,

$$\frac{G_1}{G_0} = \frac{1 - \sqrt{1-P^2}}{1 + \sqrt{1-P^2}}, \quad \frac{G_{MR}}{G_0} = \frac{P}{1 + \sqrt{1-P^2}}, \quad (12)$$

where the common prefactors $G_Q \sum_n T_n$ cancel. However, this cancellation does not occur for the ratio $G_\varphi/G_0$, where we get a factor $[\sum_n \varphi_n]/[\sum_n T_n]$ that can become arbitrarily small or large depending on the spin-mixing angles and transmission probabilities. Thus, $G_\varphi/G_0$ can for the purpose of comparing with experimental data be regarded as a fitting parameter.

**Spin-active tunneling interfaces (2nd order in $\varphi_n$ and $T_n$).** To 2nd order in the transmission probabilities and spin-mixing angles, the interfacial matrix current may be written:

$$\begin{aligned} 2\hat{I} &= G_0 \left[ \hat{g}_L, \hat{F}(\hat{g}_R) \right] + \frac{1}{4} G_2 \hat{F}(\hat{g}_R) \hat{g}_L \hat{F}(\hat{g}_R) - iG_\varphi \left[ \hat{g}_L, \hat{m}_L \right] + \frac{1}{4} G_{\varphi 2} \left[ \hat{g}_L, \hat{m}_L \hat{g}_L \hat{m}_L \right] \\ &+ \frac{i}{4} G_{\chi L} \left[ \hat{g}_L, \hat{F}(\hat{g}_R) \hat{g}_L \hat{m}_L + \hat{m}_L \hat{g}_L \hat{F}(\hat{g}_R) \right] + \frac{i}{4} G_{\chi R} \left[ \hat{g}_L, \hat{F}(\hat{g}_R \hat{m}_R \hat{g}_R - \hat{m}_R) \right] \end{aligned}, \quad (13)$$

where the matrix function $\hat{F}(\hat{g})$ is the contents of the commutator in the 1st order boundary conditions divided by $G_0$:

$$\hat{F}(\hat{g}) = \hat{g} + \frac{P}{1 + \sqrt{1-P^2}} \{\hat{m}, \hat{g}\} + \frac{1 - \sqrt{1-P^2}}{1 + \sqrt{1-P^2}} \hat{m} \hat{g} \hat{m}. \quad (14)$$

In other words, *the 2nd order boundary conditions may be written concisely as a function of the 1st order boundary conditions*. This is a new result compared to ref. 13 where the 2nd order contribution was originally derived, substantially simplifying and speeding up the numerical implementation of these boundary conditions and the solution of the Usadel equation utilizing them. We use the notation $\hat{m}$ for the magnetization experienced by transmitted particles, and $\hat{m}_L$ and $\hat{m}_R$ for particles reflected on the left and right sides of the interface, respectively. As for the new conductances that appear above, these are defined as[13]

$$G_2 = G_Q \sum_{n=1}^{N} T_n^2 \left(1 + \sqrt{1-P^2}\right)^2, \quad G_\chi = G_Q \sum_{n=1}^{N} T_n \varphi_n \left(1 + \sqrt{1-P^2}\right), \quad G_{\varphi 2} = G_Q \sum_{n=1}^{N} 2\varphi_n^2, \quad (15)$$

These conductances can be connected through $G_\chi/G_0 \cong G_{\varphi 2}/G_\varphi \cong G_\varphi G_2/2G_0^2$ if we can assume that the mean spin-mixing angle and transmission probability $\langle T \rangle$ are much smaller than their standard deviations $\Delta \varphi$ and $\Delta T$. Furthermore, it can be shown that $G_\chi/G_0 \cong \langle \varphi \rangle$; since we need less than approximately to be able to stop at a 2nd order expansion in $\varphi$, we should therefore assume that $G_\chi/G_0 < 0.3$. Finally, note that there are two different $G_\chi$ in the boundary condition: one $G_{\chi L}$ for the left side of the interface, and one $G_{\chi R}$ for the right side of the interface. For the rest of this paper, we will assume that these two conductances are equal. With all of these assumptions, we are left with a single new parameter $G_\chi$ to include in our model.

To derive the equations above, one may start with the 2nd order boundary conditions in ref. 13, and make the approximations of (i) channel-diagonal scattering $T_{nn'} = T_n$, and (ii) channel-independent polarization $P_n = P$. We will not show the derivation itself here, as the derivation is relatively straight-forward but quite lengthy.

**Spin-active reflecting interfaces (all orders in $\varphi_n$).** For a completely reflecting spin-active interface, the matrix current for arbitrarily large spin-mixing angles $\varphi_n$ can be written[13]

$$\begin{aligned} \hat{I} &= -G_Q \sum_{n=1}^{N} \left[ 1 - \frac{i}{4} \sin(\varphi_n)(\hat{g}\hat{m}\hat{g} - \hat{m}) + \frac{1}{2} \sin^2(\varphi_n/2)(\hat{g}\hat{m}\hat{g}\hat{m} - 1) \right]^{-1} \\ &\times \left[ -i \sin(\varphi_n)(\hat{m}\hat{g} - \hat{g}\hat{m}) + \sin^2(\varphi_n/2)(\hat{m}\hat{g}\hat{m}\hat{g} - \hat{g}\hat{m}\hat{g}\hat{m}) \right] \\ &\times \left[ 1 - \frac{i}{4} \sin(\varphi_n)(\hat{g}\hat{m}\hat{g} - \hat{m}) + \frac{1}{2} \sin^2(\varphi_n/2)(\hat{m}\hat{g}\hat{m}\hat{g} - 1) \right]^{-1} \end{aligned} \quad (16)$$

where $N$ is the number of scattering channels at the interface. To leading order in the spin-mixing angles $\varphi_n$, the second bracket $[-i \sin(\varphi_n)(\hat{m}\hat{g} - \hat{g}\hat{m}) + \cdots] \to -i(\varphi_n)(\hat{m}\hat{g} - \hat{g}\hat{m})$, while the first and third brackets $[1 - \cdots]^{-1} \to 1$, so the equation for the matrix current linearizes to $\hat{I} = iG_Q \sum_n \varphi_n [\hat{m}, \hat{g}]$. For comparison, the spin-mixing term in eq. (10) has the form $2\hat{I} = -iG_\varphi [\hat{g}, \hat{m}]$, and eq. (11) specifies that $G_\varphi = 2G_Q \sum_n \varphi_n$, so this can be written $\hat{I} = -iG_Q \sum_n \varphi_n [\hat{g}, \hat{m}]$. Thus, we see that the eqs (10) and (16) converge in the combined limit of zero transmission $T_n \to 0$ and weak spin-mixing $\varphi_n \ll 1$.

For simplicity, we will assume that all scattering channels have the same spin-mixing angle $\varphi$, so that $\sum_{n=1}^{N} \mapsto N$ in the equation above. Such an approximation is *e.g.* justified when there is a strong Fermi vector mismatch between the superconductor and ferromagnetic insulator[15]. The above equation is formulated at the left side of an interface; the corresponding equation at the other side of the interface is found by dropping the initial minus-sign. Using the normalization conditions $\underline{\hat{m}}^2 = \underline{\hat{g}}^2 = 1$, it is also possible to reformulate the equation above in the more economical form

$$\underline{\hat{I}} = -NG_Q\left[1 - \frac{i}{4}\sin(\varphi)\underline{\hat{a}} + \frac{1}{2}\sin^2(\varphi/2)\underline{\hat{a}}\underline{\hat{m}}\right]^{-1}$$
$$\times[-i\sin(\varphi)\underline{\hat{g}}\,\underline{\hat{a}} + \sin^2(\varphi/2)[\underline{\hat{m}},\underline{\hat{a}}]]$$
$$\times\left[1 - \frac{i}{4}\sin(\varphi)\underline{\hat{a}} + \frac{1}{2}\sin^2(\varphi/2)\underline{\hat{m}}\underline{\hat{a}}\right]^{-1} \quad (17)$$

where we have defined the auxiliary matrix $\underline{\hat{a}} = \underline{\hat{g}}\,\underline{\hat{m}}\,\underline{\hat{g}} - \underline{\hat{m}}$. Using this form of the equation, it is possible to reduce the number of matrix multiplications from 18 to 5 by reusing matrix products, which results in a more efficient numerical implementation.

**Strongly polarized ferromagnets.** In general, the propagator $\hat{g}$ has a $2 \times 2$ matrix structure in both Nambu space and spin space. For normal metals and singlet superconductors, the spin structure of the normal component is diagonal, while the spin structure of the anomalous component is antidiagonal. Explicitly written out in matrix form, this means that these materials have propagators with the $4 \times 4$ structure

$$\hat{g} = \begin{pmatrix} g_{\uparrow\uparrow} & 0 & 0 & f_{\uparrow\downarrow} \\ 0 & g_{\downarrow\downarrow} & f_{\downarrow\uparrow} & 0 \\ 0 & -\tilde{f}_{\uparrow\downarrow} & -\tilde{g}_{\uparrow\uparrow} & 0 \\ -\tilde{f}_{\downarrow\uparrow} & 0 & 0 & -\tilde{g}_{\downarrow\downarrow} \end{pmatrix}. \quad (18)$$

On the other hand, in the presence of magnetic elements and spin-dependent scattering, we also need to account for triplet superconductivity and spin-flip processes in materials, and this forces us to use the most general $4 \times 4$ form for the propagator,

$$\hat{g} = \begin{pmatrix} g_{\uparrow\uparrow} & g_{\uparrow\downarrow} & f_{\uparrow\uparrow} & f_{\uparrow\downarrow} \\ g_{\downarrow\uparrow} & g_{\downarrow\downarrow} & f_{\downarrow\uparrow} & f_{\downarrow\downarrow} \\ -\tilde{f}_{\uparrow\uparrow} & -\tilde{f}_{\uparrow\downarrow} & -\tilde{g}_{\uparrow\uparrow} & -\tilde{g}_{\uparrow\downarrow} \\ -\tilde{f}_{\downarrow\uparrow} & -\tilde{f}_{\downarrow\downarrow} & -\tilde{g}_{\downarrow\uparrow} & -\tilde{g}_{\downarrow\downarrow} \end{pmatrix}. \quad (19)$$

However, for the case of very strong ferromagnets, the spin-splitting of the energy bands can be so severe that there is effectively no interaction between quasiparticles from different spin bands. The spin structure of the propagator will then become diagonal,

$$\hat{g} = \begin{pmatrix} g_{\uparrow\uparrow} & 0 & f_{\uparrow\uparrow} & 0 \\ 0 & g_{\downarrow\downarrow} & 0 & f_{\downarrow\downarrow} \\ -\tilde{f}_{\uparrow\uparrow} & 0 & -\tilde{g}_{\uparrow\uparrow} & 0 \\ 0 & -\tilde{f}_{\downarrow\downarrow} & 0 & -\tilde{g}_{\downarrow\downarrow} \end{pmatrix}, \quad (20)$$

which means the only kind of superconductivity possible will be spin-triplet ($f_{\uparrow\uparrow}$ and $f_{\downarrow\downarrow}$). Since the propagator is diagonal in spin space for such materials, its components can also be represented as simply two decoupled propagators in Nambu space,

$$\hat{g}_{\uparrow\uparrow} = \begin{pmatrix} g_{\uparrow\uparrow} & f_{\uparrow\uparrow} \\ -\tilde{f}_{\uparrow\uparrow} & -\tilde{g}_{\uparrow\uparrow} \end{pmatrix}, \quad (21)$$

$$\hat{g}_{\downarrow\downarrow} = \begin{pmatrix} g_{\downarrow\downarrow} & f_{\downarrow\downarrow} \\ -\tilde{f}_{\downarrow\downarrow} & -\tilde{g}_{\downarrow\downarrow} \end{pmatrix}. \quad (22)$$

If we assume that the two spin-bands in the ferromagnet individually behave as normal metals, it should be reasonable to assume that the two sets of quasiparticles follow two separate metallic diffusion equations. Introducing the spin-dependent diffusion constants $D_\uparrow$ and $D_\downarrow$, these diffusion equations may be written

$$iD_\uparrow \partial_z\left(\hat{g}_{\uparrow\uparrow}\partial_z\hat{g}_{\uparrow\uparrow}\right) = \left[(\epsilon + i\eta)\hat{\tau}_3, \hat{g}_{\uparrow\uparrow}\right], \quad (23)$$

$$iD_\uparrow \partial_z \left( \hat{g}_{\downarrow\downarrow} \partial_z \hat{g}_{\downarrow\downarrow} \right) = \left[ (\epsilon + i\eta)\hat{\tau}_3, \hat{g}_{\downarrow\downarrow} \right],  \quad (24)$$

We will also define the spin-independent diffusion constant $D = D_\uparrow + D_\downarrow$ and spin-polarization $\Pi = (D_\uparrow - D_\downarrow)/(D_\uparrow + D_\downarrow)$, where we note that $D_\sigma = D(1 + \Pi\sigma)/2$. By dividing each of the above equations by its polarization factor $(1 \pm \Pi)/2$, we get

$$iD\partial_z \left( \hat{g}_{\uparrow\uparrow} \partial_z \hat{g}_{\uparrow\uparrow} \right) = \left[ 2(1 + \Pi)^{-1}(\epsilon + i\eta)\hat{\tau}_3, \hat{g}_{\uparrow\uparrow} \right], \quad (25)$$

$$iD\partial_z \left( \hat{g}_{\downarrow\downarrow} \partial_z \hat{g}_{\downarrow\downarrow} \right) = \left[ 2(1 - \Pi)^{-1}(\epsilon + i\eta)\hat{\tau}_3, \hat{g}_{\downarrow\downarrow} \right], \quad (26)$$

or if we restore the matrix notation for the spin structure,

$$iD\partial_z(\underline{\hat{g}}\partial_z\underline{\hat{g}}) = [(\epsilon + i\eta)\underline{\Pi}\hat{\tau}_3, \underline{\hat{g}}], \quad (27)$$

where we have defined the polarization matrix

$$\underline{\Pi} = \begin{pmatrix} 2/(1 + \Pi) & 0 \\ 0 & 2/(1 - \Pi) \end{pmatrix}. \quad (28)$$

This equation follows the pattern in eq. (1) if we define the matrix potential $\underline{\hat{U}} = [(\epsilon + i\eta)\underline{\Pi}\hat{\tau}_3, \underline{\hat{g}}]$, which written out becomes

$$\underline{\hat{U}} = \begin{pmatrix} 0 & 4(\epsilon + i\eta)\underline{\Pi}\underline{N}\,\underline{\gamma} \\ 4(\epsilon + i\eta)\underline{\Pi}\underline{\tilde{N}}\,\underline{\tilde{\gamma}} & 0 \end{pmatrix}. \quad (29)$$

We then extract the components $\underline{U}_{11} = 0$ and $\underline{U}_{12} = 4(\epsilon + i\eta)\underline{\Pi}\underline{N}\,\underline{\gamma}$, and invoke eq. (6) to find an equation for $\partial_z^2 \underline{\gamma}$, which reads

$$iD[\partial_z^2 \underline{\gamma} + 2(\partial_z \underline{\gamma})\underline{\tilde{N}}\,\underline{\tilde{\gamma}}\,(\partial_z \underline{\gamma})] = 2(\epsilon + i\eta)\underline{\Pi}\,\underline{\gamma}. \quad (30)$$

Thus, the only difference between Riccati parametrized diffusion equation for a normal metal and a strong ferromagnet is the occurence of the polarization matrix $\underline{\Pi}$. However, it should be stressed that the above equation was derived under the assumption that the propagator $\hat{g}$ has a diagonal structure in spin space, which implies that the Riccati parameters $\underline{\gamma}$ and $\underline{\tilde{\gamma}}$ must be diagonal as well. Thus, when implementing the equation above numerically, one must ensure that the off-diagonal terms of $\underline{\gamma}$ and $\underline{\tilde{\gamma}}$ are treated as constants and not variables; deviations from this procedure could produce numerical artifacts that violate these initial assumptions. The main motivation for writing the equation for $\underline{\gamma}$ in matrix form, is that it can now be used in a boundary condition like eq. (7) at both sides of the interface, without requiring modification. Note that the interface to a strong ferromagnet is bound to be strongly magnetized, which means that we should use eqs (10) or (17) as boundary conditions. In the limit of full polarization $\Pi \rightarrow 1$, the matrix $\underline{\Pi} \rightarrow \text{diag}(1, \infty)$. The infinite element will essentially just force the condition $\gamma_{\downarrow\downarrow} = 0$ for the spin-down component, while we get a normal metallic diffusion equation for the spin-up component,

$$iD[\partial_z^2 \gamma_{\uparrow\uparrow} + 2(\partial_z \gamma_{\uparrow\uparrow})\tilde{N}_{\uparrow\uparrow}\tilde{\gamma}_{\uparrow\uparrow}(\partial_z \gamma_{\uparrow\uparrow})] = 2(\epsilon + i\eta)\gamma_{\uparrow\uparrow}. \quad (31)$$

Physically, what happens in this limit is that the spin-splitting of the energy bands is strong enough to make the spin-up band metallic and the spin-down band insulating, which results in a so-called half-metal. Thus, we are left with two different ways to model a half-metallic ferromagnet: we can either use eq. (30), and implement a strong ferromagnet with *e.g.* $\Pi = \pm 0.999$, thus taking the limit $\Pi \rightarrow 1$ numerically; or we can take the limit $\Pi \rightarrow 1$ analytically, and implement a scalar diffusion equation for $\gamma_{\uparrow\uparrow}$ like eq. (31). We chose the first approach, since the resulting code may then be reused to model strong ferromagnets as well.

## Results and Discussion

**Density of states in S/FI/N multilayers.** To begin with, we consider the DOS in a normal metal (N) connected to a superconductor (S) via a ferromagnetic insulator (FI), which becomes modified by the existence of triplet Cooper pairs. The FI in this setup is modelled as a spin-active interface with zero spatial extent, but a finite tunneling conductance $G_0$, spin-mixing conductance $G_\varphi$, and spin-polarization $P$. Assuming no spatial extent means that the FI must have a thickness comparable to atomic length scales, which is much smaller than all the superconducting length scales in the problem. In reality, the properties of the FI would of course scale with its length, which in our model would be described by choosing a smaller tunneling conductance and larger spin-polarization at the interface. The special case of $P = 0$ was considered in refs [37] and [38] where it was shown that for a critical value of $G_\varphi$, pure odd-frequency pairing was induced at the Fermi level $\epsilon = 0$. This is manifested as a large zero-energy peak in the DOS. In Fig. 1, we now show how this effect is modified when taking into account an interface polarization $P$.

For these simulations, we set the tunneling conductance to $G_0/G = 0.3$, where $G$ is the normal-state conductance of each material. If we increase $G_0$ while keeping $G_\varphi/G_0$ and $P$ fixed, we increase the magnitude of the proximity effect, but we do not change whether it is dominated by singlets or triplets. In other words, increasing $G_0$ makes the blue regions darker and the yellow regions brighter, but does not alter the shape of the plot. Note that for zero polarization, this reproduces the well-known result that a peak suddenly appears for $G_\varphi/G_0 = 1$[37], while the value of $G_\varphi$ necessary to get a zero-energy peak gradually decreases to zero as the polarization tends to one. These results suggest that the spin-dependent transmission probabilities facilitate the conversion from singlet to triplet superconducting correlations in such a fashion that smaller spin-dependent mixing angles are required for this purpose. However, spin-dependent transmissions by themselves only weaken the singlet proximity effect: if $G_\varphi = 0$, there is no generation of triplet Cooper pairs as seen in Fig. 1 (fully gapped DOS for $G_\varphi = 0$).

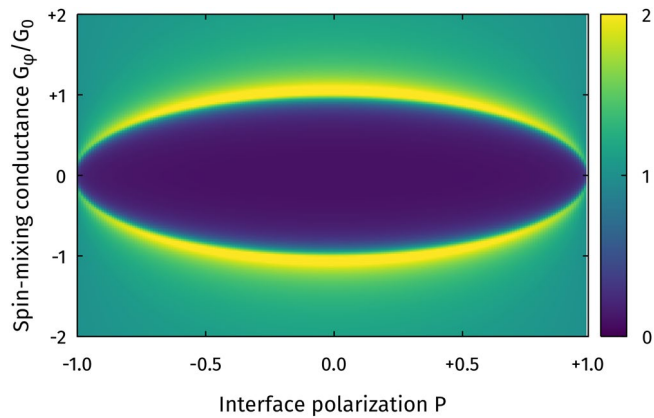

**Figure 1.** Plot of the zero-energy DOS $D(0)$ as function of interface polarization $P$ and spin-mixing conductance $G_\varphi$. The DOS was calculated in the center of the normal metal in an S/FI/N structure, where we used the BCS solution for the superconductor, treated the ferromagnetic insulator as a spin-active interface, and the normal metal was taken to have the length $\xi_S$.

In the above example, the superconductor was treated as a reservoir, meaning that the bulk propagator was used in that region. The main purpose of this paper is to determine how the superconducting region is influenced by the magnetic proximity effect, which generates triplet Cooper pairs in the superconductor. In what follows, we therefore only present self-consistent results where the superconducting order parameter and propagator are both obtained in an iterative manner. This allows us to explore how triplet Cooper pairs manifest in the superconducting region, as recently experimentally seen in refs 8 and 11.

We thus show results for a self-consistently solved DOS in both the superconducting and normal region of an S/FI/N system in Figs 2 and 3, setting the length of the superconducting region to $L_S = 3\xi_S$ and $L_S = \xi_S$ in the figures, respectively, where $\xi_S$ is the diffusive coherence length of a bulk superconductor at zero temperature. In all cases, we use the value $G_0/G = 0.3$ for the tunneling conductance. In both cases, we keep the length of the normal layer fixed at $L_N = \xi_S$. The DOS in the N region is very similar in both cases, illustrating the zero-bias peak characteristic of odd-frequency triplet pairs. It is worth to underline that although such a peak is often taken to be a signature of odd-frequency pairing, recent work has demonstrated that a system with fully gapped DOS can still exhibit strong odd-frequency pairing[39]. The DOS in the superconductor, on the other hand, changes substantially when going from $L_S = 3\xi_S$ to $L_S = \xi_S$. In the former case, the DOS only weakly deviates from the gapped bulk behavior of an s-wave superconductor. In the latter case, however, the gap is not only strongly smeared out, but a noticeable zero-energy peak emerges in the superconductor as well due to the appearance of odd-frequency triplet pairs there.

Note that in Figs 1–3, we use the definition $G_\varphi/G_0 = [\sum_n \varphi_n]/[\sum_n T_n]$, which differs by a factor $[1 + \sqrt{1-P^2}]/2$ from the definitions used in the rest of this paper. This does not change any conclusions, as this affects $G_\varphi/G_0$ by a factor 2 at most, while Fig. 1 shows that $G_\varphi/G_0$ needs to change by more than a factor 10 in order to produce a zero-energy peak at high polarizations.

It is interesting to note that one can obtain a very large zero-energy enhancement of the DOS in the superconductor, even exceeding its normal-state value, if the FI barrier itself is magnetically inhomogeneous. This is included in our model using the interfacial magnetic misalignment in the boundary conditions described previously. For a very high polarization $P$, we show how the DOS depends on the spin-mixing conductance $G_\phi$ in the left panel of Fig. 4. For large $G_\varphi$, the combination of a strongly suppressed superconducting gap $\Delta$ near the interface and the generation of triplet Cooper pairs with all spin projections (due to the interfacial magnetic misalignment) permits the DOS to completely shed its gapped character and instead develop a large zero-energy peak typical of odd-frequency pairing[40]. In the right panel, we show how the DOS develops for a fixed $G_\varphi$ when $P$ is increased, from which one infers that while a broad enhancement takes place even for $P = 0$, a sharp peak is only obtained when the spin-filtering effect of the interface is incorporated.

### Density of states in S/FI bilayers.
If the normal metal is removed, so that the superconductor is terminated by vacuum on one side and a fully reflecting magnetic insulator on the other, we have an S/FI bilayer with zero transmission of quasiparticles from S and into the FI. This can be modelled as a superconductor with boundary conditions given by eq. (17). We then find that the proximity-induced DOS in the superconductor depends strongly on both the spin-mixing angle $\varphi$ and the conductivity of the superconductor relative the number of reflective channels, parametrized by $G/NG_Q$. This is shown in Fig. 5. Moreover, the spin-mixing angle $\varphi$ strongly influences the size of the superconducting gap, as demonstrated in Fig. 6. For a thin superconductor $L_S = \xi_S$, the gap is suppressed to around 20% of its bulk value for a FI with a spin-mixing angle $\varphi/\pi = 0.9$. A larger superconductor $L_S = 3\xi_S$ is able to recover the bulk value of the order parameter at its vacuum interface, but the suppression of $\Delta$ is nevertheless substantial near the FI interface for large spin-mixing angles. The reduced gap edge is manifested in Fig. 5.

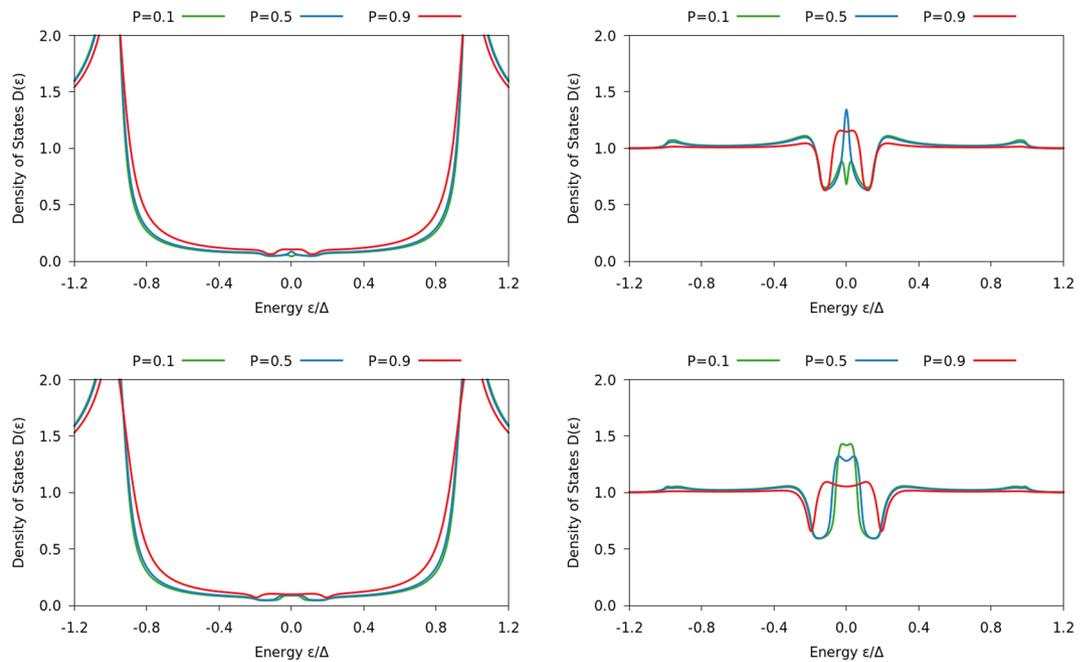

**Figure 2.** Plots of the DOS for an S/FI/N junction with $L_S = 3\xi_S$ and $L_N = \xi_S$ as function of energy. The left column shows the results on the superconducting side of the interface, and the right column on the normal-metal side. The spin-mixing conductance $G_\varphi/G_0$ is 0.75 in the top row and 1.25 in the bottom row, while the interface polarization $P$ is written in the legend.

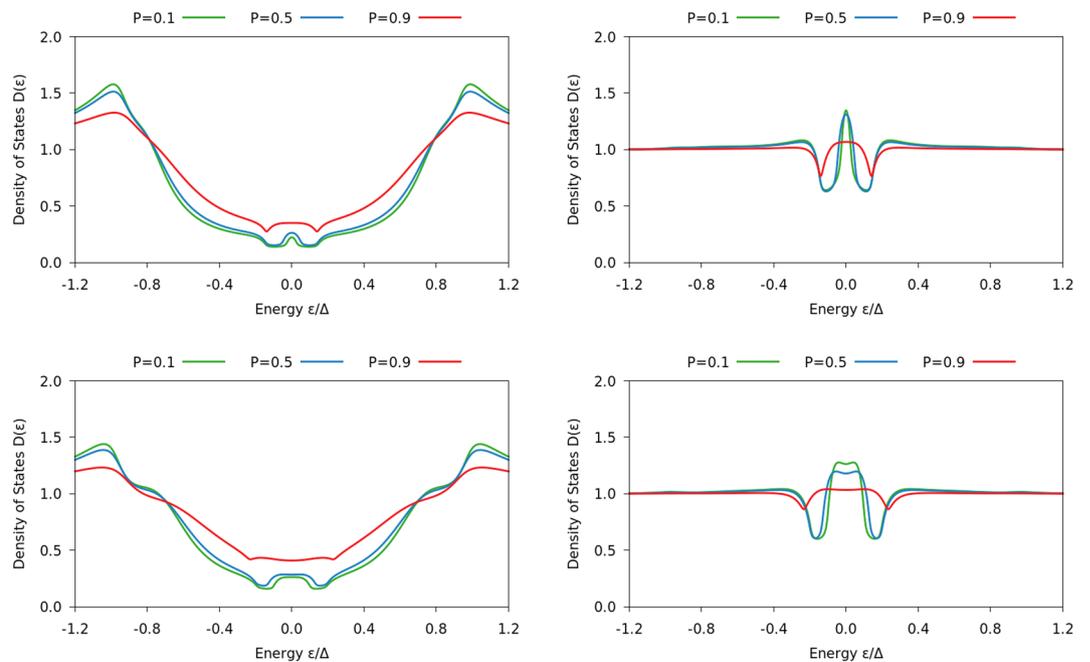

**Figure 3.** Plots of the DOS for an S/FI/N junction with $L_S = L_N = \xi_S$ as function of energy. The left column shows the results on the superconducting side of the interface, and the right column on the normal-metal side. The spin-mixing conductance $G_\varphi/G_0$ is 0.75 in the top row and 1.25 in the bottom row, while the interface polarization $P$ is written in the legend.

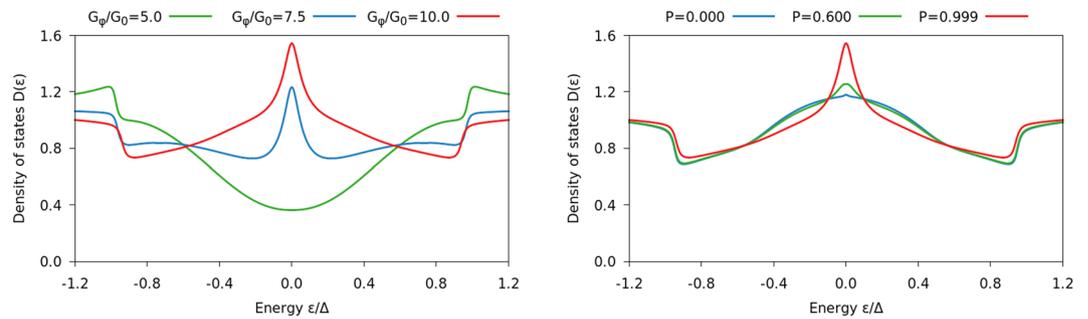

**Figure 4.** Plots of the DOS at the superconducting side of the interface in an S/FI/N junction with $L_S = 3\,\xi_S$ and $L_N = 10\,\xi_S$ as function of energy. The ferromagnetic insulator was modelled as a spin-active interface with very strong spin-mixing and polarization: in the left plot, we set the polarization $P = 0.999$ and vary $G_\varphi$, in the right plot we set $G_\varphi/G_0 = 10$ and vary $P$. In contrast to Figs 2 and 3, we also included a magnetic inhomogeneity in the model, which was incorporated by using two different magnetizations $m_L = e_x$ and $m = m_R = e_z$ in the spin-active boundary conditions.

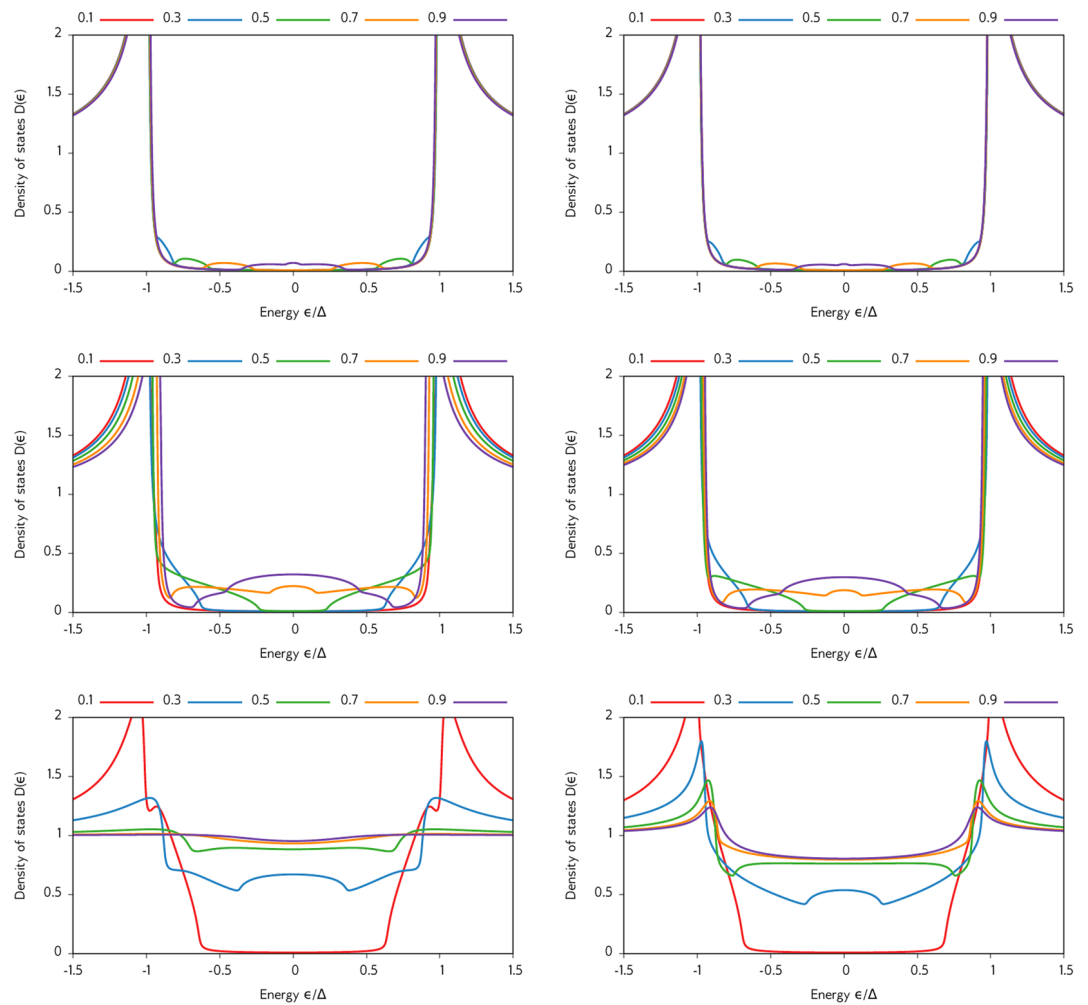

**Figure 5.** Plots of the DOS of an S/FI bilayer, measured at the superconducting side of the interface. The ferromagnetic insulator is modelled as a fully reflecting spin-active interface, and the superconductor has length $L_S = \xi_S$ in the left column and $L_S = 3\,\xi_S$ in the right one. The junction has $GL_S/NG_Q\xi_S \in \{300, 30, 3\}$, decreasing downward. The different curves correspond to different values for the spin-dependent interfacial phase shifts $\varphi/\pi$, as shown in the legends above the plots.

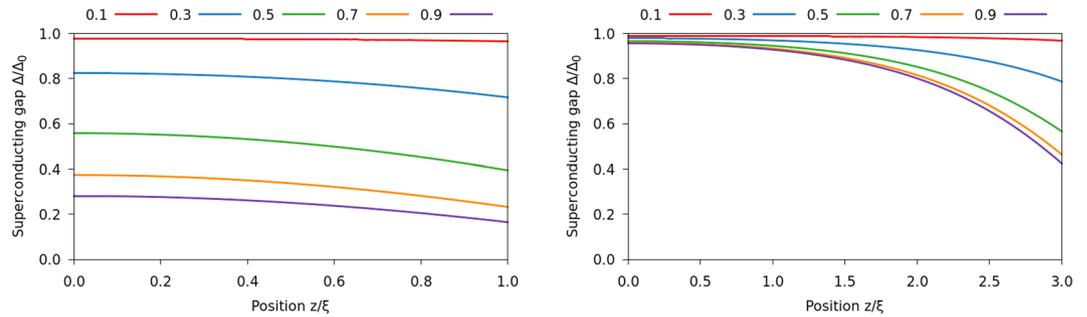

**Figure 6.** Plots of the superconducting gap in an S/FI bilayer. The superconductor has length $L_S = \xi_S$ in the left plot and $L_S = 3\,\xi_S$ in the right one. In both cases, we chose $GL_S/NG_Q\xi_S = 3$, and the spin-mixing angle $\varphi/\pi$ is shown in the legend.

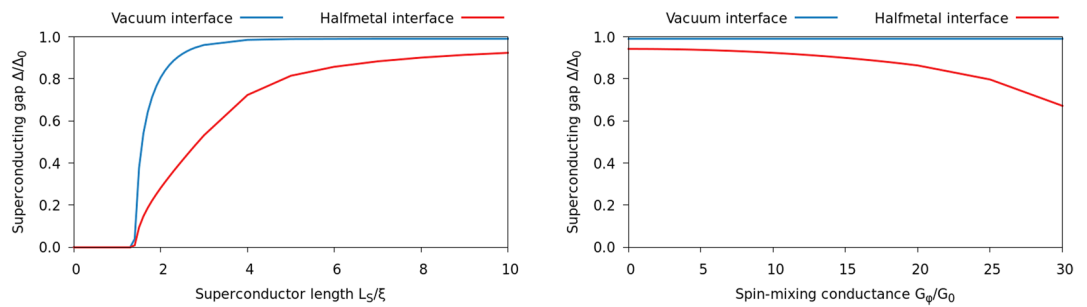

**Figure 7.** Plots of the superconducting gap at the interfaces of an S/HM bilayer. Both the plots are for a long halfmetal $L_H = 10\,\xi_S$; the difference is that we fix $G_\varphi/G_0 = 10$ but vary $L_S$ in the left plot, while we fix $L_S = 10\,\xi_S$ and vary $G_\varphi$ in the right.

We emphasize that, as shown in the Theory section, the S/FI results would be identical to the S/FI/N results in the limit of zero tunneling conductance and weak spin-mixing. However, we used a finite tunneling conductance in the previous subsection, and spin-mixing angles all the way up to $0.9\pi$ in this subsection, which is why the results are quite different.

**Density of states in S/HM bilayers.** A recent experiment by Kalcheim et al.[8] reported an unexpected result for STM-measurements on the superconducting side of a NbN/LCMO bilayer, which is precisely a S/HM structure. They found that the DOS in the superconductor could be so strongly modified by the proximity to the half-metal that all signs of gapped behavior would vanish and be replaced by a zero-energy peak that exceeded even the normal-state value. This is in stark contrast to the results we showed above for a ferromagnetic insulator, where the zero-energy peak in the superconductor always appeared inside a gapped region and whose magnitude did not exceed the normal-state value. Such a remarkably strong inverse proximity effect as seen in the experiment[8] can in fact be modelled by our theory, as we now demonstrate. For the plots below (Figs 7 and 8), we consider a S/HM bilayer and assume a $\pi/2$ magnetic misalignment at the interface. The $L_S = 3\,\xi_S$ case shown in the top left figure of Fig. 8 shows good agreement with the experimental data: a zero-bias peak which exceeds even the normal-state DOS. With increasing thickness $L_S$, a usual gapped structure is recovered. The results do, however, depend on the misalignment angle: when it is reduced to zero, the distinct zero-energy peak morphs into a weaker and more diffuse subgap plateau. This enhancement can still be larger than the normal-state DOS in some cases; e.g. when $L_S = 2\,\xi_S$, $L_H = 10\,\xi_S$, and $G_\varphi/G_0 = 10$, $D(0)$ is reduced from 1.30 with $\pi/2$ misalignment to 1.10 with no misalignment. Note also the similarity between the results in Figs 4 and 8 on the superconducting side of the interface: for similar interface parameters, we obtain nearly identical results in the S/HM and S/FI/N structures.

The generation of triplet Cooper pairs on the superconducting side has an interesting non-monotonic dependence on the length $L_S$ of the superconductor (see bottom left panel of Fig. 8), unlike the triplet proximity effect on the half-metal which decays monotonically with increasing $L_H$ (see bottom right panel of Fig. 8). For thin superconducting layers $L_S \cong \xi_S$, the superconducting gap is fully suppressed at the interface, as shown in Fig. 7. As a result, the normal-state DOS $D(0) = 1$ is obtained in the superconductor near the interface as the superconducting correlations are fully suppressed there. As $L_S$ increases, a finite value of the order parameter $\Delta$ is permitted, and around $L_S \cong 3\,\xi_S$ the largest triplet proximity effect is obtained. This is the regime where the unusually strong zero-energy peak is observed. Increasing $L_S$ even further, $D(0)$ starts to fall off rapidly and one recovers the standard BCS behavior of the superconducting DOS with a gap at low energies. We also note that as one moves away from the superconducting interface, the zero-energy peak shown in the top left figure of Fig. 8 also decreases and drops below the normal-state value $D(0) = 1$. Our theory is thus able to partially explain the experimental result of ref. 8, where the peak was observed even at the vacuum interface of the superconductor. Finally, we note that we have also solved for the DOS selfconsistently when taking into account the 2nd order boundary conditions, finding no no qualitative difference and only a very weakly suppressed magnitude of the spectral features, thus justifying the usage of the 1st order boundary conditions.

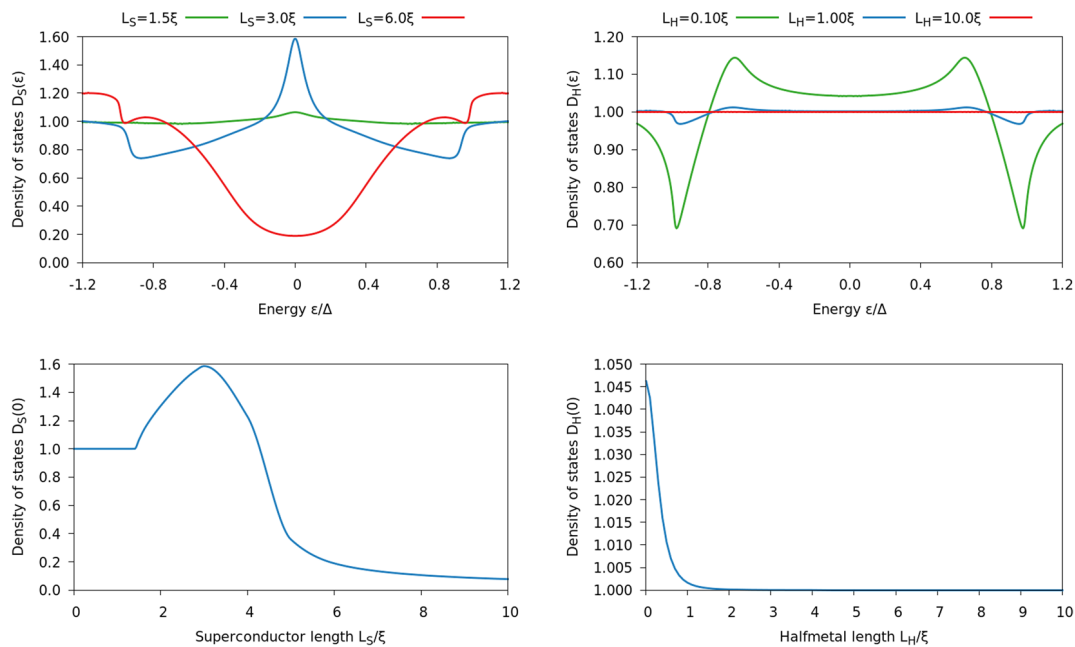

**Figure 8.** Plots of the DOS at an S/HM interface with $G_\varphi/G_0 = 10$. The left plots show how the DOS $D_S$ on the superconducting side changes with the length $L_S$ of the superconductor, when we fix the halfmetal length $L_H = 10\,\xi_S$. Conversely, the right plots show how the DOS $D_H$ on the halfmetallic side changes with $L_H$ when we set $L_H = 10\,\xi_S$. The top plots display the energy dependence of the DOS, while the bottom plots highlight the zero-energy peak.

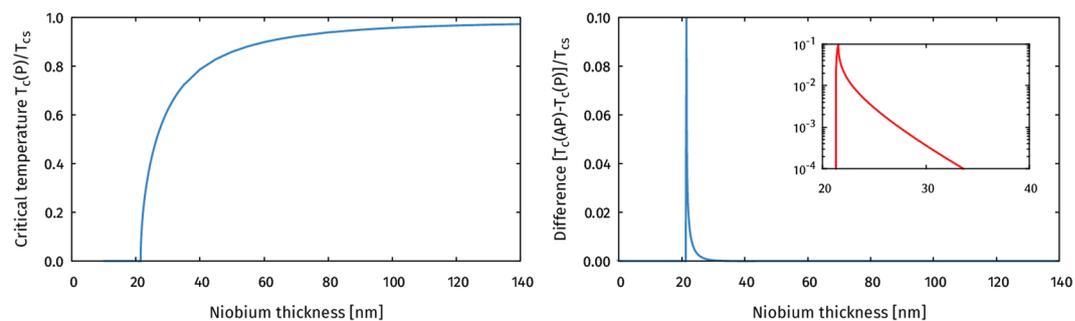

**Figure 9.** The left plot shows how the critical temperature in the parallel configuration $T_c(P)$ varies with the thickness of the superconductor. The right plot shows the critical temperature difference between the parallel and antiparallel configurations. The inset shows how this spin-valve effect decays on a logarithmic scale.

**Critical temperature in F/S/F trilayers.** Before presenting new results for the critical temperature in half-metal/superconductor hybrids, we assess how well our theory agrees with known previous spin-valve experiments. We will first consider the critical temperature of a Py/Nb/Py spin-valve structure as investigated in ref. 41. The permalloy layers were treated as homogeneous ferromagnets with an exchange field $h = 100\Delta_0$, in line with the estimate $h \cong 135$ meV for permalloy[41] and $\Delta_0 \cong 1.4$ meV for niobium. The S/F interfaces were modelled using spin-active boundary conditions with a high interface conductance $G_0/G = 1$ and spin-mixing conductance $G_\varphi/G_0 = 12$, and an experimentally realistic polarization $P = 0.38$[42]. As in the experiment, we fixed the thickness of each permalloy layer to 8 nm, varied the thickness $d$ of the niobium layer, and used a superconducting coherence length $\xi_S = 6$ nm. For each thickness of the niobium layer, we then calculated the critical temperatures $T_c(P)$ and $T_c(AP)$ for parallel and antiparallel orientations of the permalloy magnetization directions, respectively. The results are shown in Fig. 9 below.

First of all, the results show that the proximity effect in such a trilayer can be significant even for a several coherence lengths long superconductor. Superconductivity is entirely suppressed until the superconductor thickness $d \geq 21.5$ nm $\cong 3.6\, \xi_S$. After that, the critical temperature converges towards the bulk value $T_{cs}$, but is reduced by more than 1% compared to this value even for $d = 140$ nm $\cong 23\, \xi_S$. Both these results are quantitatively consistent with the results reported by Moraru *et al.* in ref. 41, where they found no superconductivity for $d < 20.5$ nm, and the critical temperature curve closely matches Fig. 9.

The right panel of Fig. 9 shows how the spin-valve effect $T_c(\text{AP}) - T_c(\text{P})$ in the system varies with the superconductor thickness. Using a critical temperature $T_{cs} = 9.2$ K for niobium, we see that the spin-valve effect abruptly rises from 0 to 0.9 K when $d = 21.3$ nm, *i.e.* the thickness at which $T_c(\text{AP})$ becomes nonzero. However, the spin-valve effect decays exponentially fast as $d$ is increased; it drops to about 0.1 K for $d = 22.7$ nm, and decreases below 1 mK for $d = 33.5$ nm. For comparison, Moraru *et al.* observed a lower spin-valve effect of about 20 mK for their best sample[41]. However, combining observations in their Figs 1 and 3, they find that the spin-valve effect drops below 1 mK for $d \geq 33$ nm, which fits very well with our results. The discrepancy in the spin-valve amplitude could *e.g.* be explained by wrong estimates for the interface parameters, or by the experimental difficulty manufacturing an ideal sample given the sensitivity of the spin-valve effect to the niobium thickness. The key observation here is nevertheless that the *spin-valve effect is completely absent for a large range of thicknesses where an inverse proximity effect exists*, *i.e.* the regime $d \geq 33$ nm in the plot.

The results for the proximity effect are remarkably robust: even if we use extreme values such as $G_\varphi/G_0 = 500$ for the spin-mixing (while keeping the other parameters as above), or a giant tunneling conductance $G_0/G = 100$ (with non-magnetic boundary conditions), the critical superconductor thickness required for $T_c \geq 0$ remains in the region 20–24 nm. This can be explained as follows. Once the properties of the interfaces become sufficiently extreme that they force the gap $\Delta \to 0$ there, then making the interface properties even more extreme cannot further suppress the gap at the interface. The strength of the spin-valve effect, on the other hand, remains more sensitive to the values of $G_\varphi$ and $G_0$.

The most notable conclusion one can draw from these results, is the extreme discrepancy that can exist between having a significant proximity effect and spin-valve effect. While the former remains visible for superconductors that are longer than 20 coherence lengths, the latter becomes negligible after just 6 coherence lengths. This is for an F/S/F spin-valve setup; for an S/F/F setup one would expect both these length scales to be reduced by at least a factor two, due to the reduced coupling between the superconductor and the second ferromagnet in the structure.

**Critical temperature in S/HM bilayers.** Inspired by the experiment by Keizer *et al.*[43], we wanted to check how an interfacial magnetic misalignment affects the critical temperature of an S/HM bilayer. This was modelled by setting $\boldsymbol{m} = \boldsymbol{e}_z$ and $\boldsymbol{m}_L = \cos\alpha\,\boldsymbol{e}_z + \sin\alpha\,\boldsymbol{e}_x$ in eq. (10); *i.e.* $\boldsymbol{m}$ was oriented along the magnetization of the half-metal, while $\boldsymbol{m}_L$ differs from it by an angle $\alpha$. First, we assumed that the superconductor was 0.7–1.5 $\xi_S$ long, that the half-metal was 12 $\xi_S$ long, that the conductance ratio was $G_0/G = 0.4$, and varied the spin-mixing conductance $G_\varphi/G_0$ in the range 0–12. Then, we fixed the length of the superconductor to $\xi_S$, and investigated the effect of varying the length of the half-metal, and the effect of including the 2nd order contributions in the boundary conditions. For each set of parameters described above, we computed the critical temperature $T_c$ for the interfacial magnetic misalignments $\alpha = 0$ and $\alpha = \pi/2$, and calculated the difference $T_c(0) - T_c(\pi/2)$ between these results as a measure of the critical temperature shift due to magnetic misalignments. The results are shown in Fig. 10.

Several noteworthy features appear. Consider first the difference between $T_c$ in the parallel and perpendicular alignment shown in the top left panel. The perpendicular configuration $T_c(\pi/2)$ is always smaller than $T_c(0)$. This can be explained physically by the fact that in the perpendicular configuration, the long-ranged proximity effect channel is opened up, allowing Cooper pairs to be converted into triplets with spin-polarization along the magnetization of the halfmetallic region and thus leak out of the superconductor. The panel also shows that for a given length $L_S$ of the superconductor, the range of spin-mixing conductance $G_\varphi$ where the device can work as a superconducting switch [$T_c(0)$ finite while $T_c(\pi/2) = 0$] is quite narrow. This is shown explicitly for a fixed length $L_S$ in the top right panel of Fig. 10. Thicker superconducting layers $L_S$ require larger spin-mixing conductance $G_\varphi$ in order to obtain the switching effect. This is physically reasonable since a larger inverse proximity effect, and thus $G_\varphi$, is required to alter the $T_c$ as the superconductor becomes bigger and acts more as a reservoir.

It is also interesting to determine how the difference in $T_c$ between the parallel and perpendicular configurations of the interface and bulk moments depend on the length $L_H$ of the half-metallic region. This is shown in the bottom left panel, where we have plotted $[T_c(0) - T_c(\pi/2)]/T_{cs}$ vs. both $L_H$ and the spin-mixing conductance $G_\varphi$. The first thing to notice is that upper horizontal line, denoting the value of $G_\varphi$ where $T_c(0) \to 0$, is completely independent on $L_H$. This is understood physically by the fact that in the parallel alignment, there is no superconducting proximity effect in the half-metal. Thus, the critical temperature of the superconductor is determined uniquely by the inverse proximity effect generated by the full reflection taking place at the interface which naturally does not depend on $L_H$.

A more surprising feature is the fact that as $L_H$ is reduced, a smaller and smaller spin-mixing conductance $G_\varphi$ is required to suppress superconductivity in the perpendicular configuration, *i.e.* $T_c(\pi/2) \to 0$. For a fixed value of $G_\varphi$, one might expect that $T_c$ is suppressed more the larger the half-metal thickness $L_H$ is. The fact that this does not occur can be explained physically as follows. For large $L_H$, the half-metal behaves essentially as a normal metal with a very weak superconducting proximity effect. This fact is corroborated by *e.g.* the behavior of the DOS in the upper right panel of Fig. 8. As $L_H$ is reduced, however, the half-metal starts to act more and more like a triplet superconductor since the only types of Cooper pairs that can exist in the half-metal are odd-frequency triplets. In other words, there is no singlet proximity effect at all in the half-metal, unlike the case in *e.g.* S/N bilayers. The key point is that the triplet superconductivity behavior is more harmful toward the host superconductor than the normal metal behavior, because in the former case there is not only a suppression of Cooper pairs but additionally a conversion from singlets to triplets. This reduces $T_c$ even further compared to when the half-metal acts as an effective one spin-band normal metal in the limit $L_H \gg \xi_S$. As a result, steadily smaller are $G_\varphi$ required to suppress $T_c(\pi/2)$ as $L_H$ is reduced.

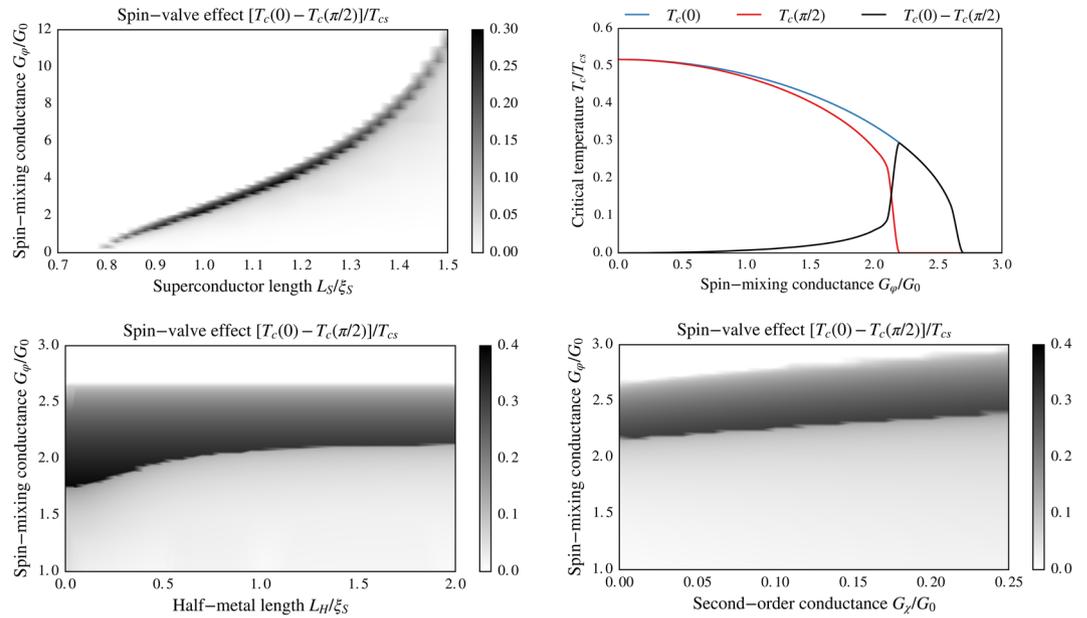

**Figure 10.** Plots of $[T_c(0) - T_c(\pi/2)]/T_{cs}$, where $T_c(\alpha)$ is the critical temperature of an S/HM bilayer with an interfacial magnetic misalignment $\alpha$, and $T_{cs}$ is the critical temperature of a bulk superconductor. Top left: We fixed the halfmetal length to $12\,\xi_S$, and varied the superconductor length and spin-mixing conductance. Above the black region, i.e. for small superconductors or strong spin-mixing, we see both $T_c(\pi/2)$ and $T_c(0)$ go to zero. Below the black region, i.e. for large superconductors and weak spin-mixing, both $T_c(\pi/2)$ and $T_c(0)$ converge to the same finite value. The black curve delineates a critical region where $T_c(\pi/2)$ drops to zero while $T_c(0)$ remains finite, leading to a very large difference. Top right: We fixed the halfmetal length to $12\,\xi_S$ and the superconductor length to $1\,\xi_S$, and highlight how $T_c(0)$ and $T_c(\pi/2)$ behave. This illustrates why the top-left curve looks like it does. Bottom left: We fixed the superconductor length to $1\,\xi_S$, and varied the halfmetal length and spin-mixing conductance. We also checked lengths $L_H$ up to $12\,\xi_S$ and find that the halfmetal length is essentially irrelevant for $L_H > 2\,\xi_S$. Bottom right: We fixed the superconductor length to $1\,\xi_S$ and halfmetal length to $12\,\xi_S$, and varied the 2nd order conductance $G_\chi$ and spin-mixing conductance $G_\varphi$. We see that the 2nd order terms basically produce a quantitative shift of the transition region towards higher values of $G_\varphi$, but does not appear to qualitatively change anything.

Finally, we have also determined the influence of including the 2nd order boundary conditions in the calculation of $T_c$. This is shown in the lower right panel of Fig. 10, revealing that there is only a small quantitative correction to the value of $G_\varphi$ providing the superconducting transition by including these additional terms parametrized by $G_\chi$. The conclusion that 2nd order terms have the same effect as a quantitative shift in $G_\varphi$ was also corroborated by DOS calculations for this setup (not shown).

**Critical temperature in S/F/N/HM multilayers.** Motivated by the recent experiment by Singh et al.[7], we have calculated $T_c$ for a superconductor/ferromagnet/normal-metal/half-metal multilayer. In accordance with the experiment, we set the superconductor thickness to $10\,\xi_S$, the half-metal thickness to $20\,\xi_S$, set the ferromagnet thickness to $0.3\,\xi_S$, and set the normal metal thickness to $1.0\,\xi_S$. For the ferromagnet, we used an exchange field of magnitude $h = 50\Delta$ in the bulk (essentially as large as quasiclassical theory permits to model the relatively strong exchange field of Ni), and set the polarization $P = 0.20$ and spin-mixing $G_\varphi/G_0 = 0.5$ at its interfaces. The superconductor used in the experiment (MoGe) had an extremely short mean free path $\ell \ll \xi_s$, firmly placing it in the diffusive limit of transport as modelled here. For the halfmetallic interfaces, we used a polarization $P = 0.999$ and spin-mixing $G_\varphi/G_0$ in the range 0–10. At all interfaces, we chose a relatively large ratio between the barrier and bulk conductances $G_0/G = 0.4$. We then calculated the critical temperature $T_c(\alpha)$, where $\alpha$ is the angle between the magnetizations of the ferromagnet and the half-metal, and used this to calculate the critical temperature shift $T_c(0) - T_c(\pi/2)$.

The result was zero critical temperature shift (with a precision of 0.0002 in $T_c/T_{cs}$). In fact, we find that both the critical temperature $T_c(0)$ with no magnetic inhomogeneity, and $T_c(\pi/2)$ with maximum magnetic noncollinearity, are essentially equal to the bulk critical temperature $T_{cs}$ for a $10\,\xi_S$ long superconductor. We therefore tried to reduce the superconductor size to below $1.0\,\xi_S$ in order to check whether that would help. In this case, both $T_c(0)$ and $T_c(\pi/2)$ were significantly reduced compared to the bulk critical temperature, with the result $T_c/T_{cs} \cong 0.7$. However, the values of $T_c(0)$ and $T_c(\pi/2)$ still ended up being equal, so that we could not find any appreciable spin-valve effect $T_c(0) - T_c(\pi/2)$. The lack of spin-valve effect indicates that the only proximity effect we find numerically is caused by the regular ferromagnet, with the halfmetallic layer being inconsequential. We therefore tried to remove the halfmetal from the system entirely, and redo the calculations for a similar superconductor/

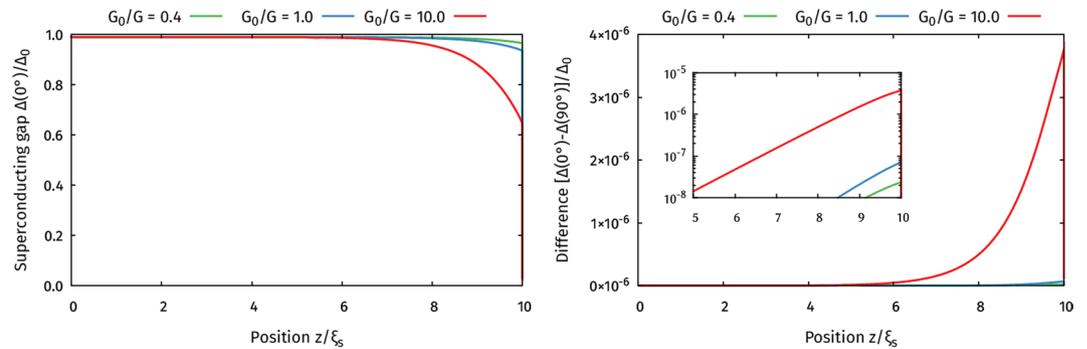

**Figure 11.** Plot of the superconducting gap $\Delta$ as a function of position inside the superconductor in the S/F/N/HM setup. The left plot shows the result for a 0° misalignment between the magnetizations of the F and HM, which is a measure of the proximity effect in the system. The right plot shows the difference between the results for 0° and 90° misalignment, which is a measure of the spin-valve effect. The inset shows how this spin-valve effect decays on a logarithmic scale. As indicated in the legends, we repeated the calculations for various interface conductances $G_0$.

ferromagnet/normal-metal multilayer, and got precisely the same critical temperature results. This indicates that the inverse proximity effect on the superconductor was dominated by the ferromagnet and not the half-metal, and that the Cooper pairs leaking from the superconductor and into the ferromagnet likely substantially decay before even reaching the normal metal. We tried checking some different lengths for the superconductor and ferromagnet, and different strengths for the spin-mixing. The highest critical temperature shift we found was for a superconductor length $L_S = 0.7\,\xi_S$ and ferromagnet length $L_F = 0.1\,\xi_S$, using $G_\varphi/G_0 = 10$ for the halfmetal interface. But even in that case, the critical temperature shift was only $[T_c(0) - T_c(\pi/2)]/T_{cs} = 0.001$. In other words, the largest simulation result we managed to achieve is two orders of magnitude smaller than the experimental result by Singh *et al.*, even after reducing the superconductor length by a factor 14 relative to the experiment, and tweaking the ferromagnet length as much as possible while remaining within the quasiclassical limits. It should however be noted that the present theory does not permit inclusion of highly transparent interfaces, in contrast to *e.g.* Nazarov's boundary conditions for non-magnetic interfaces[28], but is restricted to the limit of tunneling interfaces. For further details about our modelling of the experiment by Singh *et al.*, as well as a quantitative comparison of $T_c(0)$ and $T_c(\pi/2)$ for both S/F/N/HM and S/F/N systems with various parameters, the reader may consult the Supplementary Information.

Our results for the S/HM bilayer also show that even if the superconducting gap is strongly suppressed at the interface, it still recovers a few coherence lengths away from the interface. To investigate whether the same happens in the S/F/N/HM setup, we also performed zero-temperature calculations of the superconducting gap in this structure, as shown in Fig. 11. In all cases, we found that the proximity effect remains significant only for the first 2–3 coherence lengths away from the magnetic interface, while the spin-valve effect is insignificant for all positions and conductances. These results further support our hypothesis that the standard long-ranged proximity effect interpretation cannot fully explain the results of Singh *et al.*[7].

Previous works have considered the critical temperature of S/HM layers in the diffusive[44] and ballistic[24] limit, but cannot be compared to the measurements by Singh *et al.*[7] since these works considered a thin superconducting layer with size $L_S$ comparable to or smaller than the superconducting coherence length $\xi_S$ rather than $L_S = 10\,\xi_S \gg \xi_S$ as in the experiment. Note that in contrast to ref. 44, where it was assumed that all interfaces in the junction were transparent, we used a finite interface transparency at each interface of the S/F/N/HM junction, as this should be experimentally more realistic, and we also chose a larger magnitude of the exchange field. Since there are three such interfaces between S and HM in the junction, our structure has a much lower net transparency than in ref. 44, so that the $T_c$ variation in our case is small even for very thin superconductors $L_S < T_c$. This may explain why the $T_c$ results herein were much weaker than the one found in ref. 44 when $L_S < \xi_S$.

## Conclusion

Summarizing, we have developed a framework for studying the interaction between diffusive superconducting and strongly polarized magnetic materials and half-metals using quasiclassical theory. We have applied this framework on superconductors interfaced to strongly polarized ferromagnetic insulators and half-metallic ferromagnets, solving the equations selfconsistently in order to study the manifestation of triplet Cooper pairs induced in the superconductor. We have computed the density of states and critical temperature in the abovementioned systems. Recent experimental work have measured precisely these quantities in via STM in S/HM bilayers (DOS)[8] and resistance measurements in S/F/N/HM layers ($T_c$)[7]. We have shown that our theory is able to reproduce an unusually strong zero-energy peak in the S/HM bilayer, exceeding the normal-state value, induced in a superconductor as seen experimentally in ref. 8. We also predicted a strong spin-valve effect in such bilayers, as shown in Fig. 10. Moreover, we computed the $T_c$ shift upon 90° rotation of the magnetization in a spin-valve S/F/N/HM multilayer and discussed this result in the context of the experiment of ref. 7 and clarified the difference in length-scale for the inverse proximity effect in a superconductor and the length-scale for which a spin-valve effect occurs.

## Acknowledgements

J. Aarts, A. Singh, J. W. A. Robinson, A. Di Bernardo, and N. Banerjee are thanked for useful discussions. J.L was supported by the Research Council of Norway, Grant No. 216700 and the *Outstanding Academic Fellows* programme at NTNU. J.L. and J.A.O. were supported by the Research Council of Norway, Grant No. 240806. M.E. and M.G.B. was supported by EPSRC grant EP/N017242/1.

## Author Contributions

J.A.O. performed the analytical and numerical calculations with minor support from J.L. The majority of the writing of the manuscript was done by J.A.O. and J.L. M.E. contributed to discussion and understanding of the boundary conditions for strongly spin-polarized diffusive systems on which the manuscript is based. All authors (J.A.O., A.P, M.B., M.E., and J.L.) contributed to the discussion of the results and revisions of the manuscript.

## Additional Information

**Supplementary information** accompanies this paper at doi:10.1038/s41598-017-01330-1

**Competing Interests:** The authors declare that they have no competing interests.

**Publisher's note:** Springer Nature remains neutral with regard to jurisdictional claims in published maps and institutional affiliations.